\documentclass[12pt]{iopart}

\usepackage[british]{babel}
\usepackage{graphicx}
\usepackage{color}
\usepackage{iopams}
\usepackage{setstack}
\usepackage[T1]{fontenc}

\begin{document}

\title[Fractional Brownian motion in a finite interval]{Fractional Brownian motion
in a finite interval: correlations effect depletion or accretion zones of particles
near boundaries}
\author{T Guggenberger$^{\sharp,\dagger}$, G Pagnini$^{\dagger}$, T Vojta$^{\ddagger}$
and R Metzler$^{\sharp}$}
\address{$\sharp$ Institute of Physics and Astronomy, University of Potsdam,
14476 Potsdam-Golm, Germany\\
$\dagger$ BCAM - Basque Centre for Applied Mathematics, 48009 Bilbao, Basque
Country, Spain and Ikerbasque - Basque Foundation for Science, 48013 Bilbao,
Basque Country, Spain\\
$\ddagger$ Department of Physics, Missouri University of Science and Technology,
Rolla, MO 65409, USA}
\eads{\mailto{rmetzler@uni-potsdam.de}}

\begin{abstract}
Fractional Brownian motion is a Gaussian stochastic process with stationary,
long-time correlated increments and is frequently used to model anomalous
diffusion processes. We study numerically fractional Brownian motion confined
to a finite interval with reflecting boundary conditions. The probability
density function of this reflected fractional Brownian motion at long times
converges to a stationary distribution showing distinct deviations from the
fully flat distribution of amplitude $1/L$ in an interval of length $L$ found
for reflected normal Brownian motion. While for superdiffusion, corresponding
to a mean squared displacement $\langle X^2(t)\rangle\simeq t^{\alpha}$ with
$1<\alpha<2$, the probability density function is lowered in the centre of the
interval and rises towards the boundaries, for subdiffusion ($0<\alpha<1$) this
behaviour is reversed and the particle density is depleted close to the
boundaries. The mean squared displacement in these cases at long times converges
to a stationary value, which is, remarkably, monotonically increasing with the
anomalous diffusion exponent $\alpha$. Our a priori surprising results may have
interesting consequences for the application of fractional Brownian motion for
processes such as molecule or tracer diffusion in the confined of living
biological cells or organelles, or other viscoelastic environments such as
dense liquids in microfluidic chambers.
\end{abstract}

\noindent{\it Keywords}: anomalous diffusion, fractional Brownian motion,
reflecting boundary conditions, finite interval

\section{Introduction}
\label{sec:Introduction}

Diffusive transport is quite ubiquitous, ranging from quantum processes such as
laser cooling over thermally activated transport in living biological cells, to
the dispersal of tracer chemicals in geophysical aquifers. While theoretical
works on diffusion often consider infinite or semi-infinite domains, in many 
cases the particle motion is restricted to a finite interval. Such a scenario
is relevant, inter alia, for the dispersal of light in disordered optical
cavities, for the motion of thermally driven particles in confining microfluidic
chambers, gels, or for the molecular and (sub)micron tracer motion in the confines of
biological cells or their organelles. For normal Brownian diffusion in an interval of
length $L$ the probability density to find the particle anywhere within this interval
has the constant amplitude $1/L$ at sufficiently long times: in the stationary state
the particle can be found everywhere equally likely. This property is so engrained
in our intuition for random processes that we typically would not question its
validity. Indeed, even for generalised random processes such as the continuous
time random walk, such equidistributions naturally occur \cite{scher,bouchaud,bvp,
report}. We here show that for the widely used, Gaussian process of fractional
Brownian motion (FBM) describing overdamped viscoelastic diffusion, this property
is strikingly violated: the inherent negative or positive correlations the
stationary solution of FBM in a finite interval effect a pronounced depletion or
accretion of the probability density in the vicinity of the boundaries,
respectively.\footnote{We use the antonyms "deplete" and "accrete" in the sense
that the system studied here eventually reaches a stationary state with given
depletion or accretion zones.}

While normal diffusion is characterised by the linear time dependence $\langle
X^2(t)\rangle\propto t$ of the mean squared displacement (MSD)\footnote{Here
we restrict our discussion to the one-dimensional process $X(t)$ with initial
value $X(0)=0$. In higher dimensions the motion along different coordinates
is viewed as independent.}, \emph{anomalous diffusion\/} typically classifies
the power law time dependence
\begin{equation}
\langle X^2(t)\rangle=2K_{\alpha}t^\alpha,
\label{eq:DefAnomMSD}
\end{equation}
of the MSD, where the generalised diffusion coefficient $K_\alpha$ has physical
dimension of length$^2/$time$^\alpha$, and $\alpha$ is the anomalous diffusion
exponent. Depending on the value of $\alpha$ one distinguishes between subdiffusion
($0<\alpha<1$) and superdiffusion ($\alpha>1$) \cite{bouchaud,report}.

Anomalous diffusion has been observed experimentally and numerically in a wide
array of systems \cite{bouchaud,report}. Thus, experimentally subdiffusion is
frequently observed for passive particle motion inside living biological cells
\cite{hoefling,weiss,golding,bronshteyn,weber,weigel,tabei}
and in crowded, viscoelastic solutions \cite{weiss,wong,banks,lene1}. It was
also reported from supercomputing studies of lipid and protein molecule motion
in bilayer membranes \cite{jeon2012,kneller,gupta,prx}. Subdiffusion moreover occurs
in other systems, such as for charge carrier motion in amorphous semiconductors
\cite{scher} or chemical tracers in underground aquifers \cite{berkowitz}.
Superdiffusion has been observed for the motion of molecular motor-transported
particles inside biological cells \cite{caspi,christine,jeon_natcomm},
the motion of tracers in two-dimensional rotating flows \cite{solomon}, and for
bulk-mediated surface diffusion at liquid-solid interfaces \cite{krapf}.

Normal Brownian motion is characterised by the universal Gaussian probability density
\cite{bouchaud}. Anomalous diffusion processes lose this universality, and the various
possibilities to break the attraction of the basin of the central limit theorem give
rise to different mathematical models \cite{pccp}. For instance, the existence of
power-law sojourn times between motion events with a diverging mean waiting time, as
indeed measured in single particle tracking experiment \cite{weigel,wong},
lead to anomalous diffusion (\ref{eq:DefAnomMSD}) in the continuous time random walk
model \cite{scher,bouchaud,report,pccp}. The second very prominent anomalous stochastic
process is FBM, first introduced by Kolmogorov \cite{kolmogorov} and later studied
by Mandelbrot and van Ness \cite{mandelbrot}. FBM is a self-similar, Gaussian process
with stationary, long-time correlated increments to describe anomalous diffusion of
the power law type \eref{eq:DefAnomMSD} in the sub- and superdiffusive range $0<\alpha
<2$, see below. FBM has been identified as the governing type of motion, or an 
important ingredient of the motion, for the subdiffusion of various tracers in complex
environments both in vivo and in vitro \cite{weber,tabei,lene1,jeon2012,kneller,guigas},
but also for completely different stochastic processes such as electronic
network traffic \cite{mikosch} or financial time series \cite{comte,biagini}.
In the superdiffusive regime with $1<\alpha\le2$ positive increment
correlations and single trajectory power-spectra consistent with FBM were observed for
the actively driven motion of endogenous granules inside amoeba cells as well as for
the motion of the amoeba themselves \cite{christine,gleb}.

Concurrent to its wide use in diverse fields FBM has been studied in mathematical
literature quite extensively, see, for instance, \cite{biagini,beran,qian}.
Nevertheless, many of its fundamental properties remain elusive, especially in the
presence of non-trivial boundary conditions. Thus, the method of
images\footnote{Similar to the images method in electrodynamics
a reflecting or absorbing boundary condition at $x=0$ for Brownian motion with
initial position $x_0>0$ can be taken into account by placing a second, imaginary
particle (the "image") at the reflected point $-x_0$ and then summing up (reflecting
boundary) or subtracting (absorbing boundary) the respective Green's functions $P(x,
t)$ with $P(x,0)=\delta(x)$, such that the full solution of the boundary value problem
is $Q(x,t)=P(x-x_0,t)\pm P(x+x_0,t)$ \cite{images}.} typically
applied for Brownian motion or random walks with long-tailed waiting time
distributions \cite{images} fails and there is no known generalised diffusion
equation for FBM that could be solved with direct methods for a given
set of boundary values. Among the few results known are the (asymptotic) density
of first passage times of FBM confined in a semi-infinite domain \cite{ding,
molchan} and conjectures for a wedge domain \cite{jeon_epl}, as well as the
probability density of
FBM confined in a semi-infinite interval with an absorbing boundary at the origin
in a first order perturbation theory approach \cite{wiese}.\footnote{We note that
FBM should not be confused with scaled Brownian motion (SBM) defined in terms of a
diffusion equation with a power-law time dependent diffusivity $D(t)\simeq t^{\alpha
-1}$ with $0<\alpha<2$ \cite{lim}. SBM is Markovian and its probability density
function obeys a generalised diffusion equation that can be solved for given
boundary conditions.} The approach of choice to study FBM in the presence of
boundary conditions in most cases is therefore by simulations. We mention that
while FBM is asymptotically ergodic, that is, the time average of its MSD in the
long time limit converges to the ensemble limit \cite{deng}, it is transiently
ageing \cite{kursawe} and non-ergodic in an external confinement \cite{lene1,pre12}.
We finally note that FBM is the basis for a class of stochastic processes called
generalised grey Brownian motion \cite{gianni}, which was used to model anomalous
diffusion in biological systems \cite{gianni1}.

Motivated by a recent study of FBM in a semi-infinite interval with a reflecting
boundary at the origin \cite{vojta}, we here investigate by extensive simulations
FBM confined to a finite interval with reflecting boundary conditions. The central
result is that the naively expected constant amplitude $1/L$ in an interval of
length $L$ in the stationary limit for Brownian motion is replaced by a solution
for FBM in which the amplitude closer to the boundaries is decreased or increased 
for subdiffusive and superdiffusive FBM. We analyse this reflected FBM in terms of
the probability density function and the MSD. In section \ref{sec:Theory} we briefly
introduce FBM (section \ref{sec:Fbm}) and discuss the implementation of reflecting
boundary conditions leading to reflected FBM (section \ref{sec:ReflectedFbm}). In
section \ref{sec:Results} we present and discuss our results. We present our
Conclusions and an outlook in section \ref{sec:Conclusion}. A comparison of our
results for an alternative implementation of reflecting boundary conditions is
given in the appendix.

\section{\label{sec:Theory}A primer on fractional Brownian motion and its
numerical implementation}

\subsection{\label{sec:Fbm}Fractional Brownian motion}

FBM is a centred Gaussian process with covariance function 
\begin{equation}
\left\langle X(t_1) X(t_2)\right\rangle=K_\alpha\bigg(t_1^\alpha+t_2^\alpha-
\left|t_1-t_2\right|^\alpha \bigg)
\label{eq:covFbm}
\end{equation}
and continuous sample paths defined for anomalous diffusion exponents in the
interval $0<\alpha<2$ \cite{biagini}. For $\alpha=1$ FBM is a Wiener process
describing Brownian motion. From definition (\ref{eq:covFbm}) it follows
that FBM starts at the origin, $X(0)=0$, has the MSD \eref{eq:DefAnomMSD},
and the free-space Gaussian probability density function
\begin{equation}
P(x,t)=\frac{1}{\sqrt{4\pi K_\alpha t^\alpha}}\exp\left(-\frac{x^2}{4K_\alpha
t^\alpha}\right).
\label{eq:PdfFbm}
\end{equation}
Furthermore, the increments of FBM are stationary but---except for the case of
Brownian motion ($\alpha=1$)---long-time correlated and hence not independent.
For superdiffusion ($\alpha>1$) and subdiffusion ($\alpha<1$), respectively, the
increments are positively and negatively correlated \cite{biagini}, see below.
These correlations are the
cause for the strongly non-Markovian nature of FBM and the lack of a complete
mathematical apparatus to analytically deal with the process in the presence of
a fixed length (and, by virtue of the MSD, thus time) scale. FBM is a self-similar
process with parameter $\alpha/2$: for all $c>0$, in distribution $X(ct)=c^{
\alpha/2}X(t)$. This means that the paths of FBM are time scale-invariant up to a
constant factor, that is, they statistically appear the same in rescaled time
intervals.

Despite the deceivingly simple form of the probability density function
(\ref{eq:PdfFbm}), naively employing the method of images to construct the
first passage time density leads to erroneous results \cite{molchan,jeon_epl,
oleksii}.\footnote{Expression (\ref{eq:PdfFbm}) is in fact identical with the
probability density function of scaled Brownian motion in unbounded space for
which the generalised diffusion equation is known and Markovian. Using the
method of images based on solution (\ref{eq:PdfFbm}) to determine the survival
of a particle for an absorbing boundary is consistent with the first passage
behaviour of scaled Brownian motion.}
Passing over to a discrete time version of FBM that can be implemented numerically,
we here define reflected FBM as follows. Discrete time FBM is taken as $Y_n=
X(\epsilon n)$, where $\epsilon>0$ is a time step. Its increment process $R_n=Y_{n+1}
-Y_n$ is discrete time fractional Gaussian noise (FGN). From this definition it
follows that discrete FGN is a stationary, centred Gaussian process with covariance
\begin{equation}
\langle R_iR_{i+j}\rangle=K_\alpha\epsilon^\alpha\bigg(\left|j+1\right|^\alpha
+\left|j-1\right|^\alpha-2\left|j\right|^\alpha\bigg).
\label{eq:covarianceFgn}
\end{equation}
The random variables defined by FGN are thus identically Gaussian-distributed, but,
except for the case of Brownian motion with $\alpha=1$, long time correlated and
hence not independent, such that the emerging process FBM is strongly non-Markovian.
For superdiffusion (subdiffusion) the increments are positively (negatively)
correlated,
\begin{equation}
\label{eq:FgnCorrelationSign}
\langle R_i R_{i+j}\rangle\left\{\begin{array}{ll}>0,&\alpha>1\\
=0,&\alpha=1\\
<0,&\alpha<1\end{array}\right..
\end{equation}
In the long-time limit $j\to\infty$ the correlations tend to zero, the asymptotic
form reading $\langle R_iR_{i+j}\rangle\sim K_\alpha\epsilon^\alpha\alpha(\alpha-1)
j^{\alpha-2}$.

Using the definition of discrete time FGN, discrete time FBM satisfies the recursion
relation
\begin{equation}
\label{eq:recursionFreeFbm}
Y_0=0,\quad Y_{n+1}=Y_n+R_n
\end{equation}
with the solution
\begin{equation}
Y_n=\sum_{i=0}^{n-1} R_i.
\label{eq:solutionRecursionFreeFbm}
\end{equation}
Discrete time FBM can thus be considered as a random walk with identically
Gaussian distributed but long-time correlated steps. By simulating discrete
time FGN and using the recursion relation \eref{eq:recursionFreeFbm}, discrete
time FBM can be directly obtained.

\subsection{\label{sec:ReflectedFbm} Reflected fractional Brownian motion}

To implement the reflecting boundary condition and thus define discrete time
reflected FBM $Z_n$ we use discrete time FGN and modify the recursion relation
\eref{eq:recursionFreeFbm} appropriately. On the semi-infinite interval $[0,
\infty)$ we define reflected FBM by
\begin{equation}
\label{eq:recursionReflectedFbmSemiInfinite}
\begin{array}{ll}Z_0=0,\\Z_{n+1}=\left|Z_n+R_n\right|.\end{array}
\end{equation}
This means that if the particle attempts to jump across the boundary of the
interval by a distance $d>0$ to the left, that is, attempts to perform the move
$0\leq Z_n\to Z_n+R_n=-d<0$, it is instead reflected and placed inside the
interval with the same distance to the boundary. While definition
\eref{eq:recursionReflectedFbmSemiInfinite} is local in space and identical to
the one used for normal Brownian motion, due to the built-in correlations of
FGN, the probability density function of reflected FBM does not possess a
horizontal derivative at the boundary for $\alpha\neq1$, as was reported in
\cite{vojta}. The standard horizontal-derivative boundary condition for a reflecting
boundary is thus not valid for FBM, which is intimately connected with the
above statement that the method of images cannot be applied.

The definition of the recursion relation for discrete time reflected FBM in a
\emph{finite\/} interval $[a,b]$ ($a<0<b$) is based on the same idea for either
of the two boundaries. However, for a finite interval one has to consider the
possibility that the particle jumps across the boundary by a distance greater
than the interval length $L=b-a$, that is, $d>L$. Then, simply reflecting the
particle as for the semi-infinite interval would place it outside the interval.
While in our simulations below the interval length $L$ is much larger than the
average single jump length and thus such effects essentially never occur, to
fully rule out this possibility the particle is kept reflecting alternately
at both boundaries until it is finally reflected into (placed inside) the
interval. This procedure indeed works for any $L$ and $d$ and leads to the
recursion relation \eref{eq:recursionReflectedFbmFinite}.
Our study here demonstrates that even in the
stationary limit for $n\to\infty$ the probability density function is not
given by the constant value $1/L$, apart from the normal Brownian case without
correlations. Instead the value of the probability density is significantly
depleted or accreted for sub- and superdiffusion. This phenomenon will
effect non-negligible consequences on natural systems in which tracer
particles follow the laws of FBM, as we discuss in section \ref{sec:Conclusion}.

Note that reflected FBM could, in principle, also be defined differently. In
\ref{sec:Appendix} we demonstrate that this alternative choice does not lead
to the same behaviour and already produces inconsistent results for normal
Brownian diffusion. In the following we employ the meaningful definition of
reflected FBM given in equation (\ref{eq:solutionRecursionFreeFbm}).

\section{\label{sec:Results}Probability density function and mean squared
displacement of reflected fractional Brownian motion}

For the simulation of discrete time FGN we employed the Cholesky method
\cite{dieker}, which can be used to simulate an arbitrary Gaussian process,
given its expectation and covariance function. As underlying random number
generator we used the SIMT-oriented Fast Mersenne Twister \cite{saito}. The
anomalous diffusion coefficient was set to $K_\alpha=1/2$ in units of
cm$^2/$sec$^\alpha$ and the anomalous diffusion exponent $\alpha$ ranged
from 0.5 to 1.8. We used a symmetric interval $[-L/2,L/2]$ centred at the
origin, and the interval length $L$ was changed from 10 to 2,000 in units
of cm. The time step was set to $\epsilon=1$ in units of sec, and we performed
simulations with up to $N=2\times 10^4$ time steps and $M=5\times10^5$ trajectories,
chosen such as to guarantee sufficient convergence of the results. In all runs the
initial condition placed the particle at the origin, $X_0=0$. In the following
all quantities corresponding to unconfined FBM are indicated with the subscript
"free".

\subsection{\label{sec:PDF}Probability density function}

\begin{figure}
\centering
\includegraphics[width=8cm]{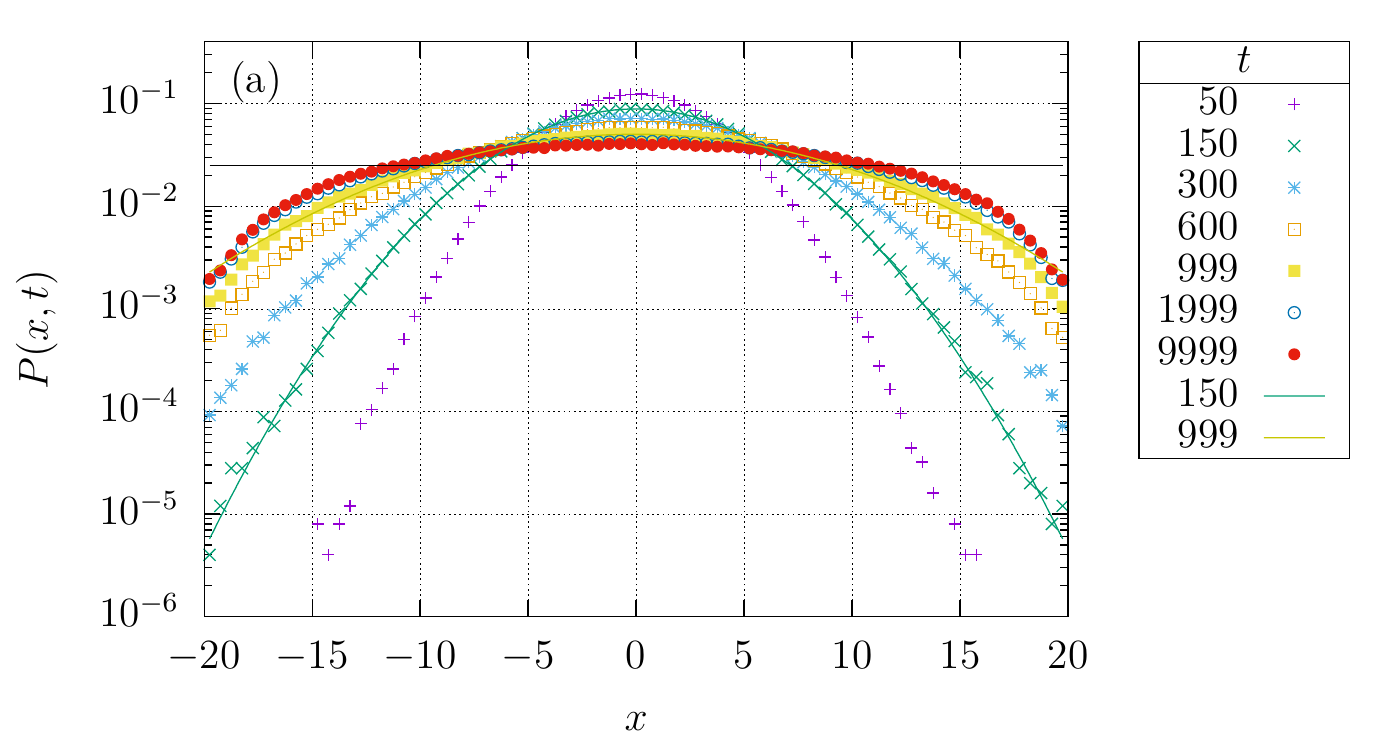}
\includegraphics[width=8cm]{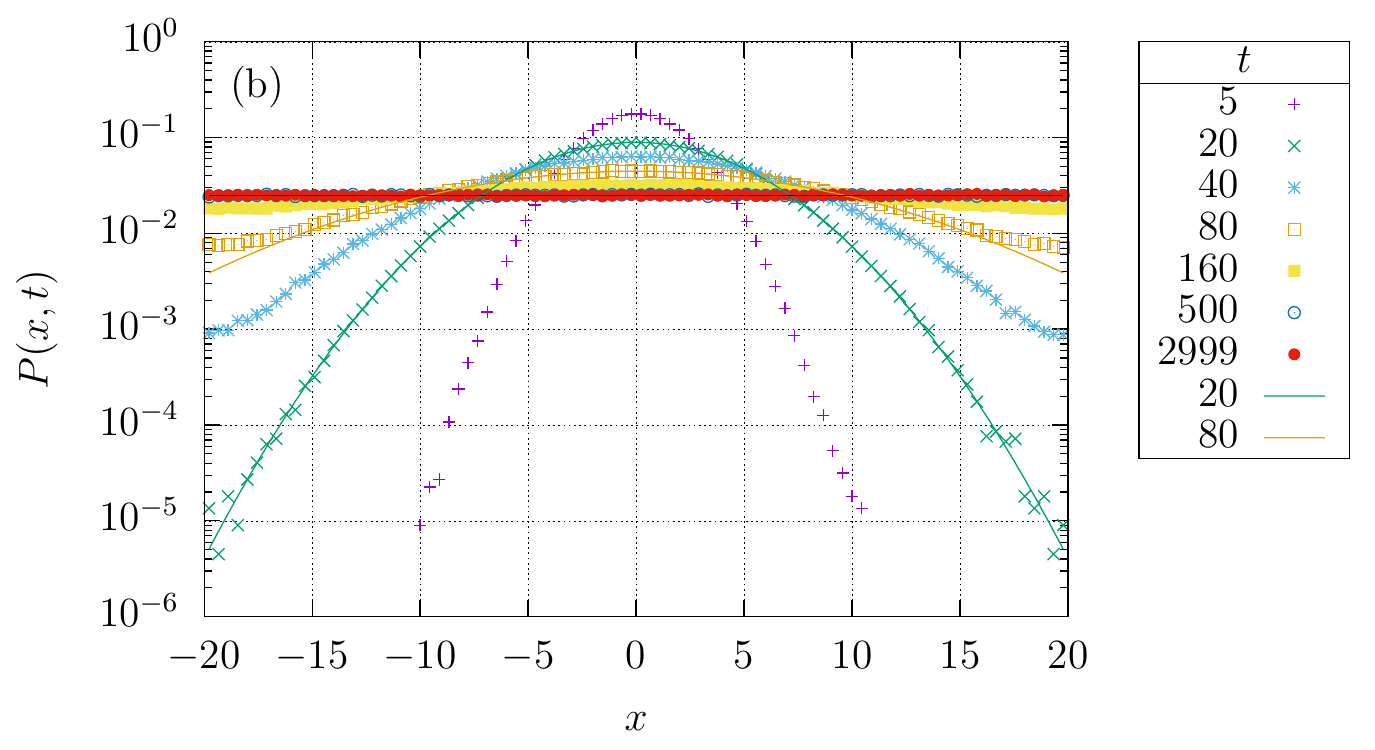}
\includegraphics[width=8cm]{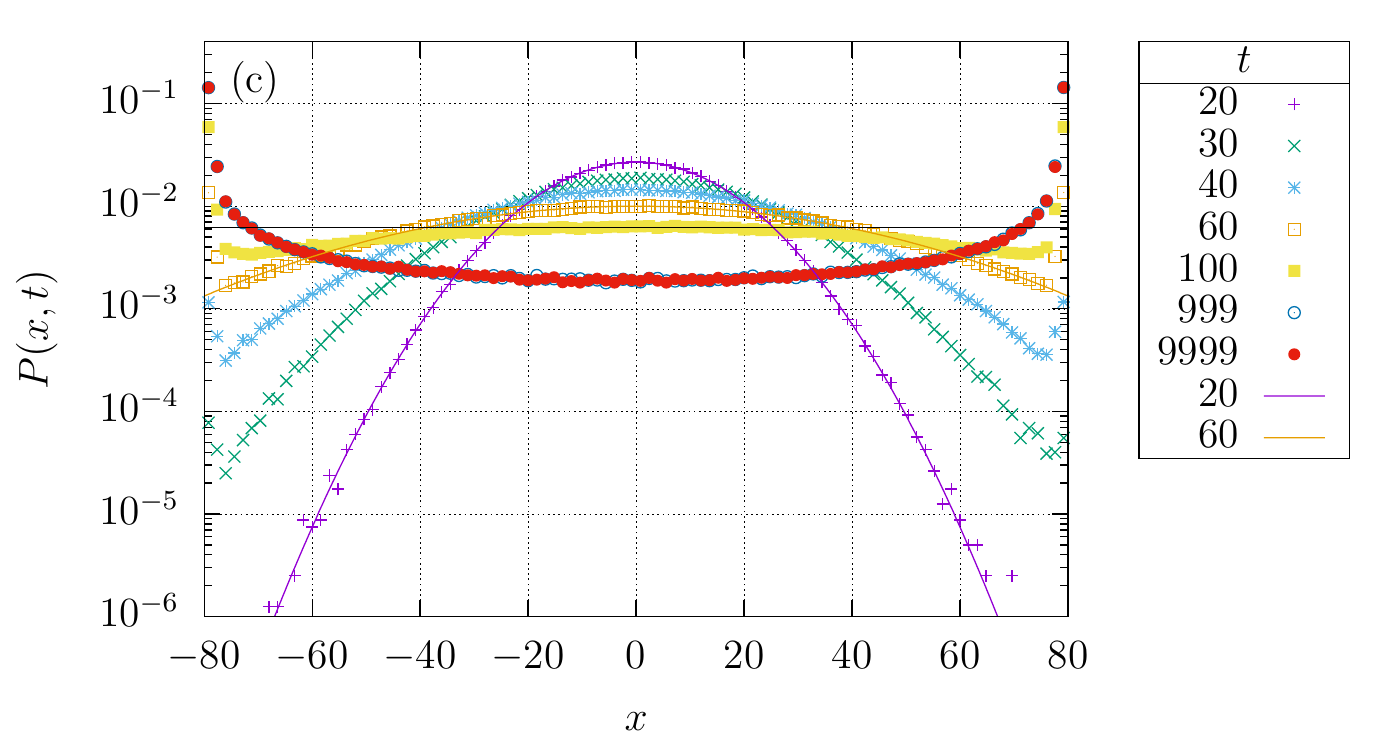}
\caption{\label{fig1} Probability density function of reflected FBM shown at
different times (see figure legend) for three different anomalous diffusion
exponents, from top to bottom: (a) subdiffusion, $\alpha=0.6$;
(b) normal Brownian diffusion, $\alpha=1$; (c) superdiffusion,
$\alpha=1.8$. The respective interval length is $L=40$,
40, and 160. The coloured lines show the theoretical probability density function
of FBM, the black line shows the equidistribution with amplitude
$1/L$ on the interval.}
\end{figure}

Figure \ref{fig1} shows the probability density function of reflected FBM for
three different anomalous diffusion exponents. We make the following observations.
(i) At all times the shape of the probability density is symmetric with respect to
the origin, $P(-x,t)=P(x,t)$, consistent with the symmetric definition of the
process. (ii) At sufficiently short times the probability density coincides with
that of the corresponding free FBM, equation \eref{eq:PdfFbm}, as it should be.
(iii) At longer times the effect of the boundary becomes significant and the
probability density deviates from that of free FBM. We note that
even in the superdiffusive case in figure \ref{fig1}c the solution of free FBM still
shows very good agreement with reflected FBM at intermediate times, apart from the
region close to the boundaries, demonstrating a relatively slow propagation of the
boundary-effected disturbance of the probability density. In the limit of very long
times the probability density function converges to a stationary limit.
(iv) Strikingly, while for normal Brownian diffusion the stationary form of the
probability density has the constant amplitude $1/L$, for sub- and superdiffusion
the stationary shape deviates significantly from this equidistribution. Namely,
compared to the $1/L$-equidistribution, for subdiffusion the stationary probability
density is increased in the central region of the interval and monotonically
decreases towards the boundary, attaining amplitudes significantly below $1/L$
over an appreciable boundary zone. In contrast, for superdiffusion the stationary
probability density is decreased in the central region of the interval and
monotonically increases towards the boundary. Moreover, for
superdiffusion at intermediate times the behaviour of the probability density is
non-monotonic between the origin and the boundary.

\begin{figure}
\centering
\includegraphics[width=8cm]{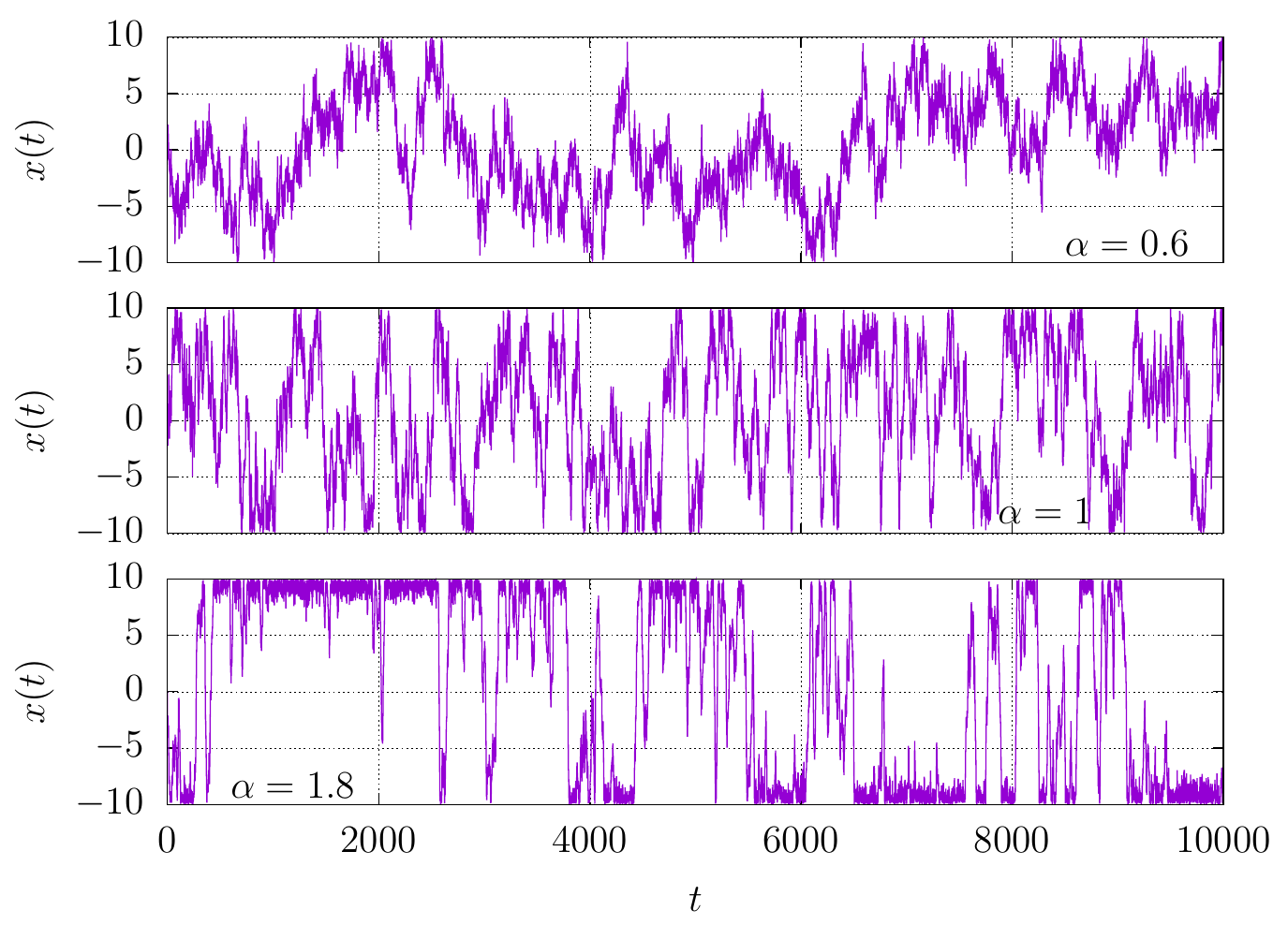}
\caption{\label{fig2} Sample paths of reflected FBM as function of time on the
interval $L=20$ for the three anomalous diffusion exponents $\alpha=0.6$, $1$,
and $1.8$ (top to bottom).}
\end{figure}

The behaviour of the probability density of reflected FBM can be explained
qualitatively as follows. For subdiffusion the jumps $R_n$ that the particle
performs are negatively correlated. Therefore, if the particle jumps across
the boundary and gets reflected in one step, in the next step it tends to
jump to the opposite direction, away from the boundary. Hence, the particle
on average tends to stay away from the boundary region and thus the probability
density in that region is depleted. For superdiffusion the jumps are positively
correlated and therefore, if the particle jumps across the boundary and gets
reflected in one step, in the next step it tends to jump in the same direction,
towards the boundary. This leads to an accretion of particle probability in
the boundary region. For normal Brownian diffusion, in contrast, the jumps are
uncorrelated and hence the particle has no tendency to stay in or stay away
from the boundary region. This effects the simple equidistribution with amplitude
$1/L$. Figure \ref{fig2} shows a typical sample path for normal Brownian diffusion,
sub- and superdiffusion, and nicely illustrates this behaviour.

\subsection{\label{sec:MSD} Mean squared displacement}

\begin{figure}
\centering
\includegraphics[width=8cm]{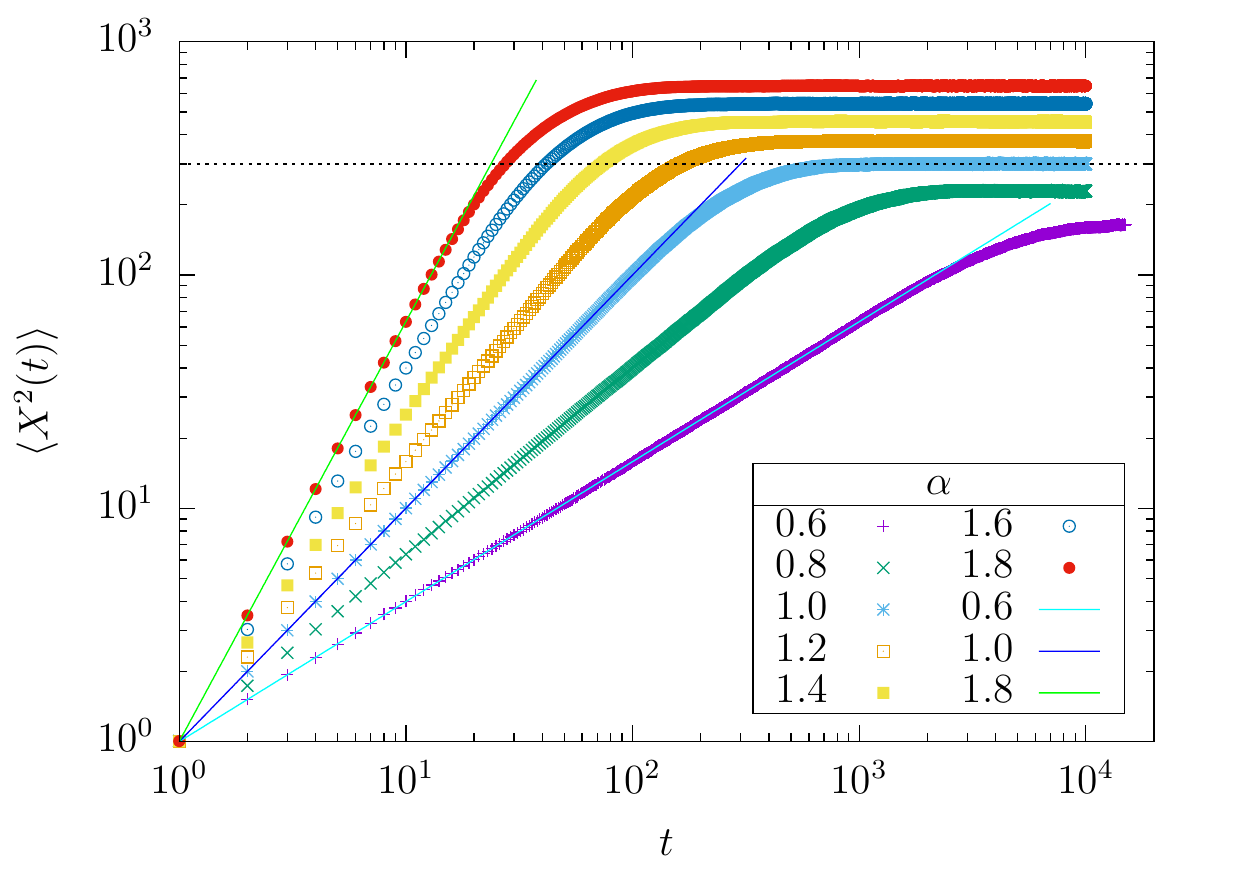}
\includegraphics[width=8cm]{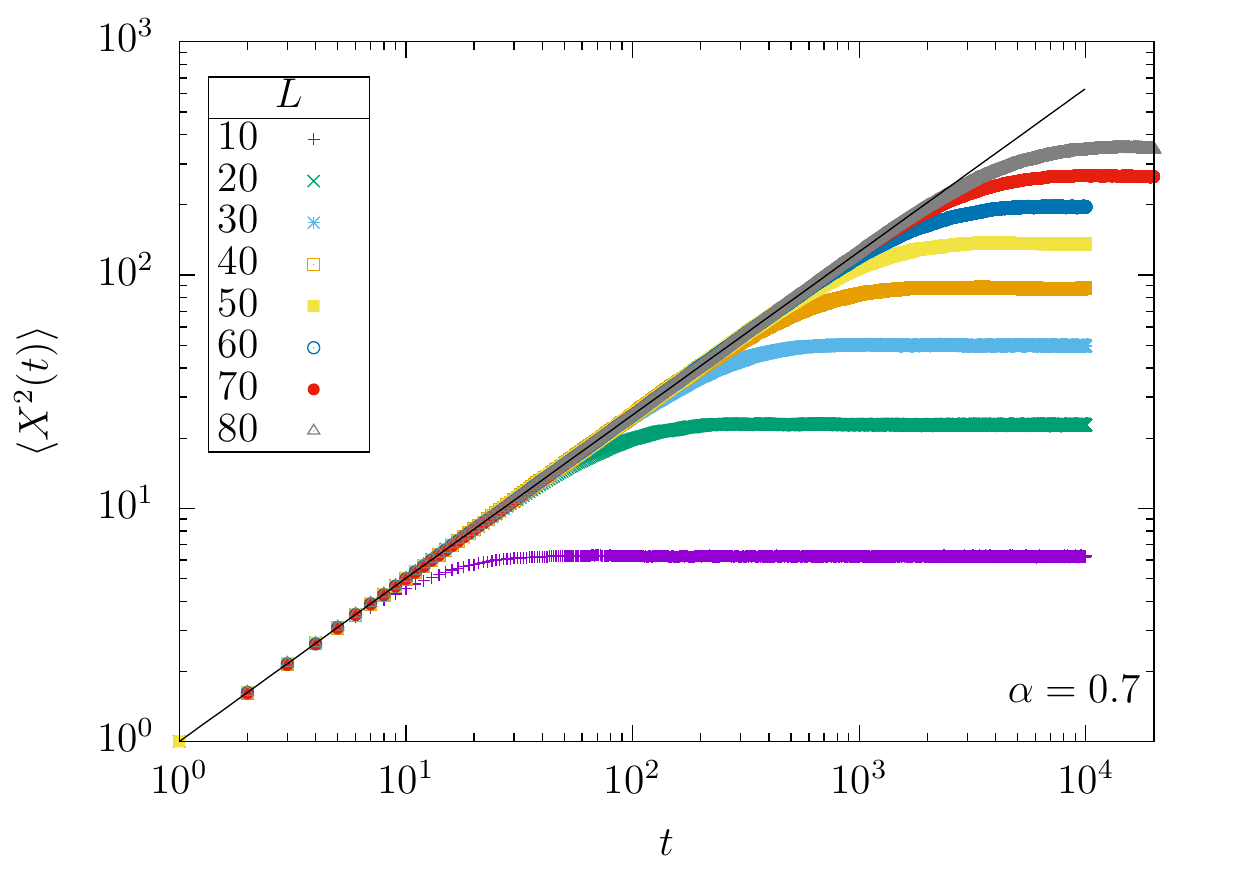}
\caption{\label{fig3} MSD of reflected FBM with interval length $L=60$ for
different anomalous diffusion exponents (top) and with anomalous diffusion
exponent $\alpha=0.7$ for different interval lengths (bottom). The solid lines
show the theoretical MSD of the corresponding free FBM \eref{eq:DefAnomMSD}. The
dashed line shows the stationary value $L^2/12$ of the MSD in the Brownian case
$\alpha=1$, see equation \eref{eq:statMSDRBM}.}
\end{figure}

We now show that the effects observed for the probability density function
also translate to the behaviour of the MSD. Figure \ref{fig3} shows the MSD
of reflected FBM with interval length $L=60$ for different anomalous diffusion
exponents (top) and with anomalous diffusion exponent $\alpha=0.7$ for different
interval lengths (bottom). At short times the MSD behaves like that of the
corresponding free FBM, as it should. However, at long times the MSD converges
to the stationary value $x_\mathrm{st}^2(\alpha,L)=\lim_{t\to\infty}\langle X^2(t)
\rangle$, which is monotonically increasing with the anomalous diffusion exponent
$\alpha$ and the interval length $L$. Since in the case of reflected Brownian
motion ($\alpha=1$) the stationary probability density is given by the
equidistribution $P_\mathrm{st}(x)=1/L$, in that case the stationary value of the
MSD simply becomes
\begin{equation}
\label{eq:statMSDRBM}
x_\mathrm{st}^2(\alpha=1,L)=\int_{-L/2}^{L/2}\frac{x^2}{L}\mathrm{d}x=\frac{L^2}{12}.
\end{equation}
This value is shown as dashed line in figure \ref{fig3}.

\begin{figure}
\centering
\includegraphics[width=8cm]{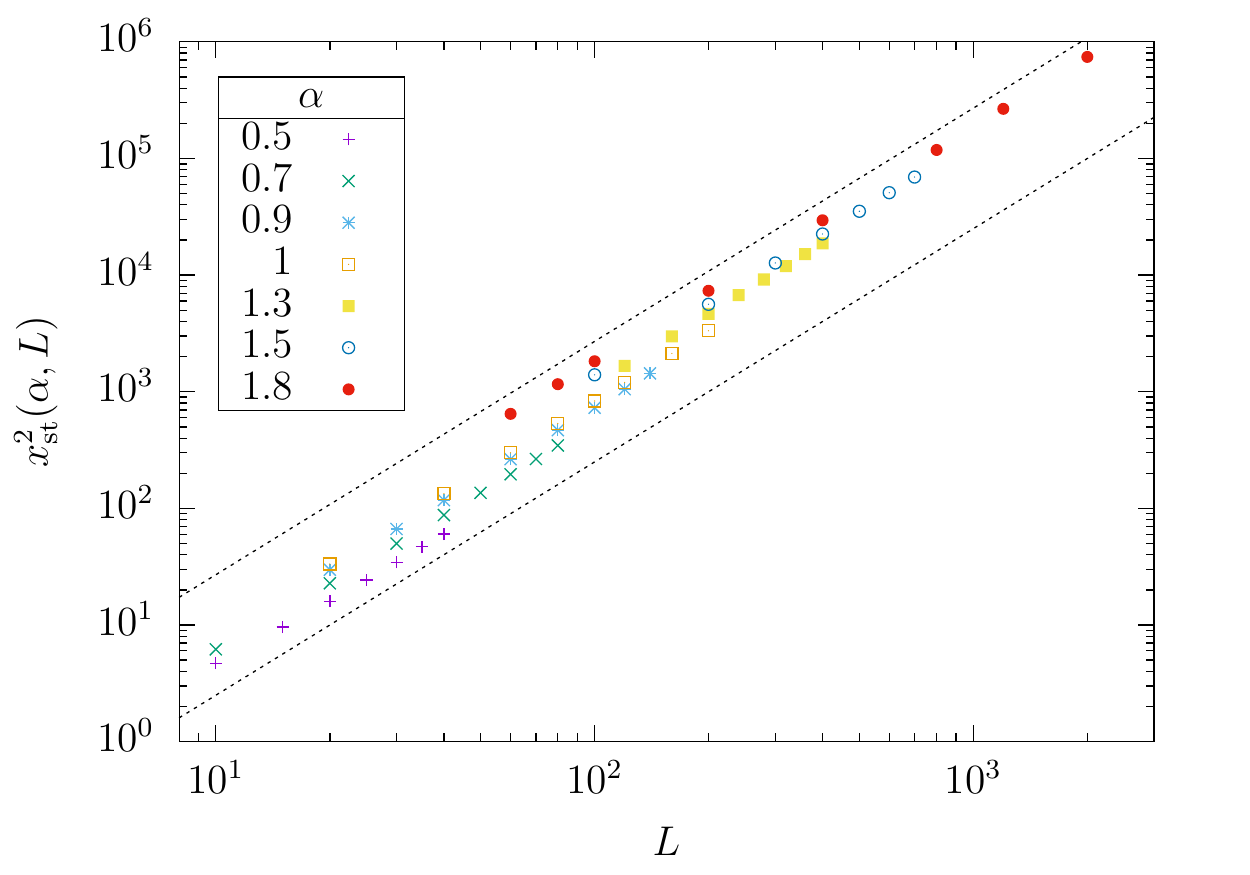}
\caption{\label{fig4} Stationary value $x^2_{\rm st}(\alpha,L)$ of the MSD 
as a function of the interval length $L$ for different anomalous diffusion
exponents $\alpha$. The dashed lines are proportional to $L^2$.}
\end{figure}

We determined the MSD of reflected FBM for a range of $\alpha$ values and
several interval lengths $L$ for each fixed $\alpha$ (the case $\alpha=0.7$ is
shown in figure \ref{fig3}). From these we determined the stationary values
of the MSD as mean values from the stationary plateaus. Figure \ref{fig4}
shows the stationary value of the MSD versus the interval length for the
different anomalous diffusion exponents. The dashed lines corroborates that
the stationary value of the MSD is proportional to the squared interval
length, independent of the anomalous diffusion exponent. We fitted\footnote{We
used the gnuplot fit routine based on the non-linear least-square
Marquardt-Levenberg algorithm.} the stationary values of the MSD for each
$\alpha$ as a function of $L$ with the fit-function $f(L)=aL^b$ and fit
parameters $a$ and $b$. The resulting values of the fit parameters shown in
table \ref{tab:fitValues} nicely corroborate the conjecture of the
$L^2$-proportionality. The slight deviation from the value 2 for the case
$\alpha=0.5$ is likely due to finite size effects, as for this most subdiffusive
value the attained value is quite short, and in figure \ref{fig3} the plateau
has not been fully reached.

\begin{table}
\centering
\begin{tabular}{|c|c|c|}\hline
$\alpha$ & $a$ & $b$ \\\hline\hline
0.5 & 0.05 & 1.91\\
0.7 & 0.06 & 1.98\\
0.9 & 0.07 & 2.00\\
1.0 & 0.08 & 2.00\\
1.3 & 0.11 & 2.01\\
1.5 & 0.14 & 2.00\\
1.8 & 0.18 & 2.00\\\hline
\end{tabular}
\caption{\label{tab:fitValues}Values of the fit parameters $a$ and $b$,
resulting from fits of the stationary value of the MSD as a function of $L$
for each $\alpha$. The statistical uncertainties of the fit-parameter values
are smaller than $10^{-2}$.}
\end{table}

From the $\alpha$ dependence of the stationary MSD shown in figure \ref{x2_vs_alpha}
for different interval lengths $L$ we deduce that the data are consistent with the
functional form $x^2_{\mathrm{st}}(\alpha,L)=\alpha^c\times L^2/12$, where $L^2/12$
is its value in the Brownian limit $\alpha=1$, and $c>0$.

\section{\label{sec:Conclusion}Conclusion}

We studied by simulations the stochastic process of reflected FBM, which is
confined to a finite interval with reflecting
boundary conditions, a situation that is typical for tracer particles in
the viscoelastic confines of biological cells or their organelles, as well
as in artificially crowded liquids in microfluidic devises and similar.
We found that the stationary probability density function of this reflected
FBM significantly deviates from the equidistribution with amplitude $1/L$
found for reflected Brownian motion ($\alpha=1$). In particular, for
superdiffusion ($1<\alpha<2$) the stationary probability density is decreased in the
centre of the interval and particle accretion occurs towards the boundaries, reaching
amplitudes above $1/L$. For subdiffusion ($0<\alpha<1$) this behaviour is reversed,
and we observe a distinct depletion of the probability density in an appreciable
region around the boundaries. The MSD at long times converges to a stationary
value, which is monotonically increasing with rising anomalous diffusion exponent
$0<\alpha<2$ and interval length $L>0$. Our simulation results corroborate that
the stationary value of the MSD is proportional to $L^2$ for all $\alpha$.

The form of the stationary probability density function can be qualitatively
explained as a result of the interplay between the positive or negative
correlations of the discrete time FGN (the steps used to generate reflected
FBM) and the boundary conditions. It is tempting to ask to which extent the
observed form of the stationary probability density function is universal for
centred Gaussian processes with stationary, positively (negatively) long-time
correlated increments under confinement with reflecting boundary conditions.
In the case of positive correlations, it is conjectured in reference \cite{vojta}
that for increments asymptotically decaying for large $n$ as $n^{\alpha-2}$ with
$1<\alpha<2$ the stationary probability density equals that of the corresponding
reflected FBM, and for a decay faster than $n^{-1}$ the stationary probability
density agrees with that of reflected Brownian motion. For short-time correlations
(positive or negative with at least exponentially fast decay) we naturally expect
the stationary probability density to agree with that of reflected Brownian motion.
A more detailed study of such processes will be of interest. Similarly, it should
be analysed how the relaxation behaviour of both the MSD and the probability
density function looks like when we introduce hard or soft cutoffs to the FGN,
as recently studied in \cite{garcia}. Finally, it will be interesting to see
how corresponding stochastic processes fuelled by FGN but with distributed
(superstatistical) diffusivities \cite{andy} behave under confinement.

As mentioned above FBM is widely used to model anomalous diffusion in various
complex systems, particularly for (sub)micron-sized tracer particles such as
vesicles, granules, viruses, or tracer beads in the crowded cytoplasm of
biological cells or in artificially crowded liquids. Inside cells, but also
in many situations in vitro, boundaries in the form of cellular membranes or
microfluidic chambers play an essential role. Cognisance of depletion layers
around boundaries in the case of passive subdiffusion or accretion layers for
actively driven, superdiffusive particles will likely affect model calculations
for interactions with the boundaries, for instance, the binding to membrane
embedded receptors. It is an interesting question how these boundary effects
conspire with other, concurrent effects. Thus, in a microfluidic chamber or
inside a simple membrane vesicle, the depletion/accretion effects due to
reflected FBM may superimpose with transient sticking to the boundary. This
could be studied quantitatively in a reflected FBM model with a sticking time
distribution to the reflecting boundary, or in terms of reactive (Robin)
boundary conditions. In real biological cells, we could think of even more 
complicated situations. Thus, many cells have an enriched layer of actin
cytoskeleton close to the cell wall. Passive tracer particles may therefore be
trapped intermittently in cages \cite{wong,aljaz}, counteracting the depletion
effects due to subdiffusive FBM. Both experiments and detailed simulations will 
be necessary to scrutinise this phenomenon. In this context we may venture a
simple scaling argument. The fact that the stationary MSD data show an $L^2$
dependence with a prefactor different from $1/12$ implys that the extra
contribution to the probability density caused by the boundaries is not confined
to a finite interval close to the walls, as otherwise a convergence to the value
$1/12$ for large $L$ should be observed. Indeed, such a situation would not be
surprising if we assume that the behaviour of the probability density is similar
to the power-law found for a semi-infinite domain in \cite{vojta}. We also note
that the $L^2$ proportionality of the stationary MSD puts a constraint on the
exact form of the probability density. If we assume naive scaling, at least in
the limit $L\gg K_{\alpha}^{1/2}$ at unit time we would expect a functional
behaviour of the stationary probability density of the form $P(x,L)=(1/L)g(x/L)$,
where $g(\cdot)$ is a scaling function. Consistent with our arguments, the
depletion/accretion zone width thus scales with $L$ and becomes a non-negligible
effect. The exact determination of the width of the depletion or accumulation
layer as function of $L$, $K_{\alpha}$ and $\alpha$ will be the topic of future
research.

\begin{figure}
\centering
\includegraphics[width=8cm]{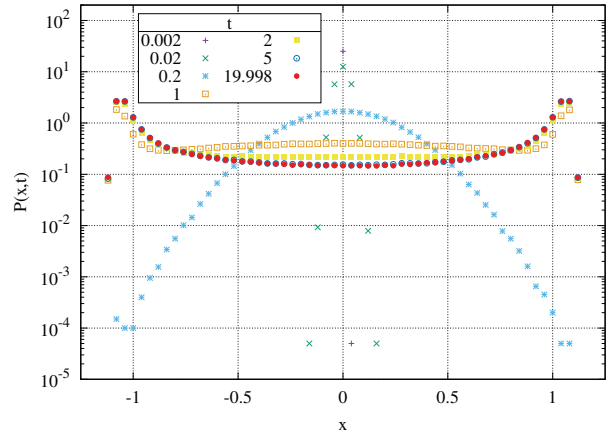}
\caption{Probability density function of superdiffusive FBM with
$\alpha=1.8$ in the external potential $V_{16}(x)=x^{32}/160$ for varying times.
Similar to reflected FBM, accretion zones emerge in the wings of the stationary
distribution.}
\label{fig5}
\end{figure}

It is instructive to compare our results for the MSD in our finite interval $L$
with those for FBM confined in an harmonic potential $V(x)=(k/2)x^2$ ($k>0$) as
studied in \cite{pre12,oleksii}. The main difference, for any stochastic process,
is that the random motion in an interval with reflecting boundaries (infinitely
steep confining potential) is athermal, that is, its stationary state does not
involve the diffusion coefficient. In a confining potential, processes such as
normal Brownian diffusion or continuous time random walks with any distribution
of waiting times always converge to the corresponding Boltzmann solution with a
well defined temperature \cite{report,pccp}. FBM, in contrast to the fractional
Langevin equation fulfilling Kubo's fluctuation-dissipation relation \cite{kubo,
goychuk}, converges to a stationary state depending on both the generalised
diffusivity $K_{\alpha}$ and the anomalous diffusion exponent $\alpha$, a fact
that follows from the corresponding (overdamped) Langevin equation fuelled by
FGN \cite{pre12,oleksii}. At short times the MSD of this process grows like that
of unconfined FBM. At long times it converges to the stationary value
$x_{\mathrm{st}}^2(k,\alpha)=(K_\alpha/k^\alpha)\Gamma(\alpha+1)$ in terms of
the $\Gamma$-function. Its stationary value, in contrast to the strictly
monotonic dependence on $\alpha$ of $x_{\mathrm{st}}^2$ for reflected FBM, is
in general a non-monotonic function of $\alpha$ \cite{oleksii}. Moreover, it can
be shown that the transition of the MSD to the
stationary value is exponential with the single characteristic time scale
$1/k$, such that stationarity is reached at a time independent of $\alpha$
\cite{pre12}. In contrast, for reflected FBM the time at which stationarity is
reached strongly depends on $\alpha$, as evidenced by figure \ref{fig3} (top).
Hence, the MSD of reflected FBM behaves fundamentally different from that of
FBM confined to move in an harmonic potential. We expect that this fundamental
differences between reflected and potential-confined FBM persists when the
harmonic potential is generalised to steeper potential $V_n(x)
=kx^{2n}/(2n)$ with $n>2$. However, what happens in the limit $n\to\infty$? In
this case, the potential essentially describes a potential well in $[-1,1]$ with
infinitely high walls. Hence, the particles can move freely inside that region,
but are still confined to it (by the infinitely high potential walls), similar
to the case with reflecting boundary conditions. This argument
is indeed corroborated by the data in figure \ref{fig5}. It remains open exactly
for which $n$ such a crossover can be observed and what the associated relaxation
dynamics is.

The case of an ever steeper external potential may also be used as an argument
in favour of the approach chosen herein with respect to the FGN driving the
reflected FBM process. Namely, we once create a time series of noise increments that
we use as input in our simulations, upon reflection at one of the two walls the
FGN time series is simply continued. One may ask whether the memory of the FGN
due to the built-in power-law correlations should not be reset once a reflection
event occurs. No such complication is expected when we consider FBM in a moderate
external potential, for instance, the above-mentioned harmonic potential. Here the
action of the potential is taken to superimpose to the FGN, consistent with the
quantative modelling of experiments \cite{lene1}. If we now make the external
potential much steeper, we would expect that the argument of un-interrupted FGN
still holds. However, as we demonstrate in figure \ref{fig5} for the superdiffusive
case already in this case significant accretion zones are created.\footnote{Note
that this effect is similar to the emergence of multimodal distributions for
strongly confined L{\'e}vy flights \cite{multimodal}.} The ultimate answer whether
the approach with invariant FGN is justified in systems such as biological cells
will, of course, have to come from experiments.

\ack TG and RM acknowledge funding through grants ME 1535/6-1 and ME 1535/7-1
of Deutsche Forschungsgemeinschaft. GP acknowledges funding from the Basque
Government through the BERC 2014-2017 and BERC 2018-2021 programmes, and from
the Spanish Ministry of Economy and Competitiveness MINECO through BCAM Severo Ochoa
excellence accreditation SEV-2013-0323 and SEV-2017-0718 as well as through project
MTM2016-76016-R
"MIP". RM acknowledges an Alexander von Humboldt Polish Honorary Research
Scholarship from the Polish Science Foundation.

\appendix

\section{\label{sec:Appendix}Alternative boundary conditions produce inconsistent
results}

Based on the procedure of alternate reflections described in section
\ref{sec:ReflectedFbm} one arrives at a recursion relation for discrete time
reflected FBM in a finite interval by the following argument. Denote with $D_n=
\min\{|Z_n+R_n-a|,|Z_n+R_n-b|\}$ the distance of the particle to the nearest
boundary after its $(n+1)$th jump. If this leads to particle across a boundary
(for instance, $Z_n+R_n>b$), $D_n$ is the distance of the particle to this
boundary. Then $S_n=D_n\ \rm{mod}\ L$ is the jump length of the last reflection,
by which the particle is finally placed inside the interval. If the interval
length $L$ "fits into" the distance $D_n$ evenly ($\lfloor D_n/L\rfloor$ even)
the last reflecting boundary is equal to the one jumped across first.\footnote{The
symbol $\lfloor\cdot\rfloor$ denotes the floor function. For any real number $x$,
$\lfloor x\rfloor$ is the largest integer less than or equal to $x$: $\lfloor x
\rfloor=\max_{n\in\mathbb{Z}}$ for $n\leq x$.} If $\lfloor D_n/L\rfloor$ is odd,
the last reflecting boundary is opposite to the one jumped across first. Hence,
we define discrete time reflected FBM in the finite interval $[a,b]$ by $Z_0=0$
and
\begin{equation}
\label{eq:recursionReflectedFbmFinite}
Z_{n+1}=\left\{\begin{array}{ll}
Z_n+R_n,&a\leq Z_n+R_n\leq b\\[0.2cm]
b-S_n,&Z_n+R_n>b\ \rm{and}\lfloor D_n/L\rfloor\ \rm{even}\ \rm{or}\\
&Z_n+R_n<a\ \rm{and}\ \lfloor D_n/L\rfloor\ \rm{odd}\\[0.2cm]
a+S_n,&Z_n+R_n>b\ \rm{and}\ \lfloor D_n/L\rfloor\ \rm{odd}\ \rm{or}\\
&Z_n+R_n<a\ \rm{and}\ \lfloor D_n/L\rfloor\ \rm{even}.
\end{array}
\right.
\end{equation}

One can easily think of other boundary conditions which can justifiably be called
"reflecting" and which thus lead to corresponding "reflected" FBM. A particular
simple choice is defined such that if the particle jumps across the boundary, it
is placed exactly on this boundary. We call this a "condensed reflecting boundary
condition". Hence, the corresponding condensed reflected FBM in the interval $[a,b]$
is defined by $Z_0=0$ and
\begin{equation}
\label{eq:recursionRFBM2}
Z_{n+1}=\left\{\begin{array}{ll}
Z_n+R_n,&a\leq Z_n+R_n\leq b\\[1pt]
a,&Z_n+R_n<a\\[1pt]
b,&Z_n+R_n>b\end{array}\right.
\end{equation}
In this appendix we compare our simulation results for reflected FBM defined by
\eref{eq:recursionReflectedFbmFinite} shown in the main text with results for
condensed reflected FBM \eref{eq:recursionRFBM2}.

\begin{figure}
\centering
\includegraphics[width=6.8cm]{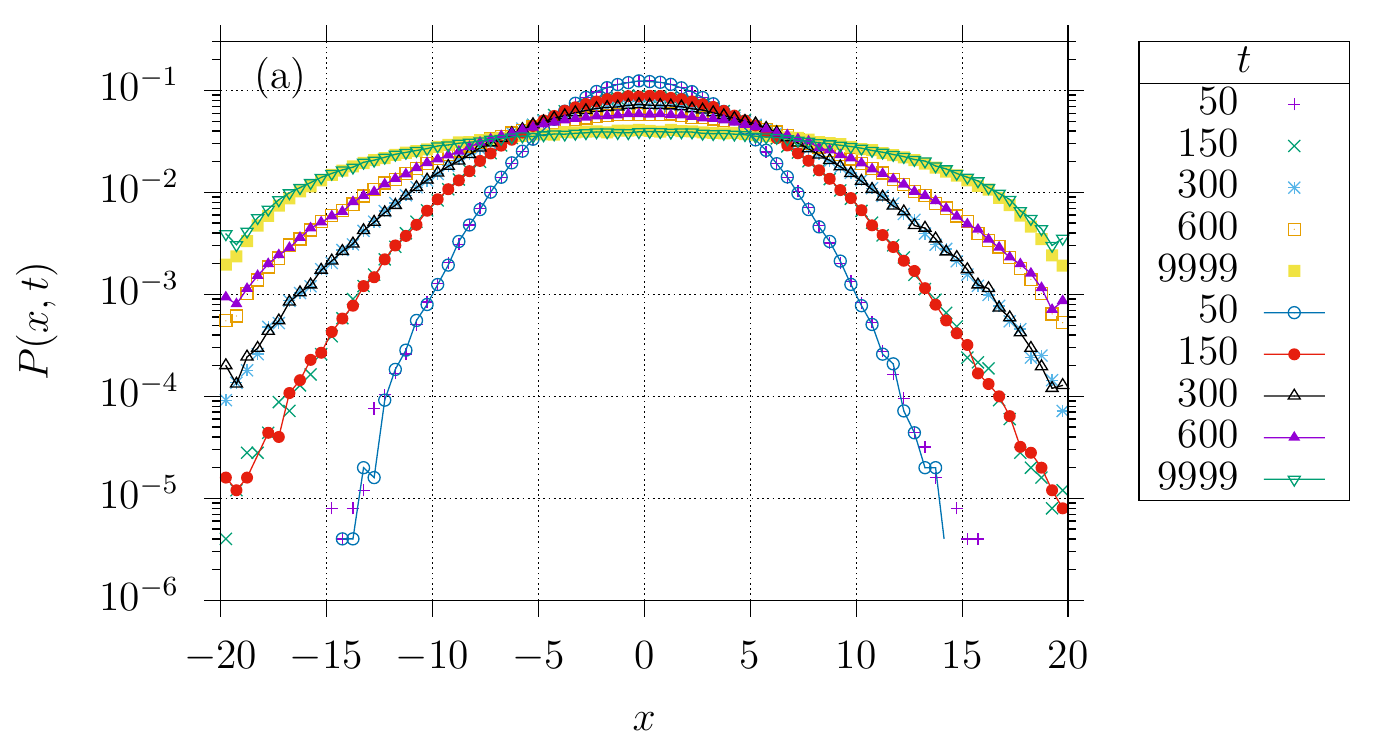}
\includegraphics[width=6.8cm]{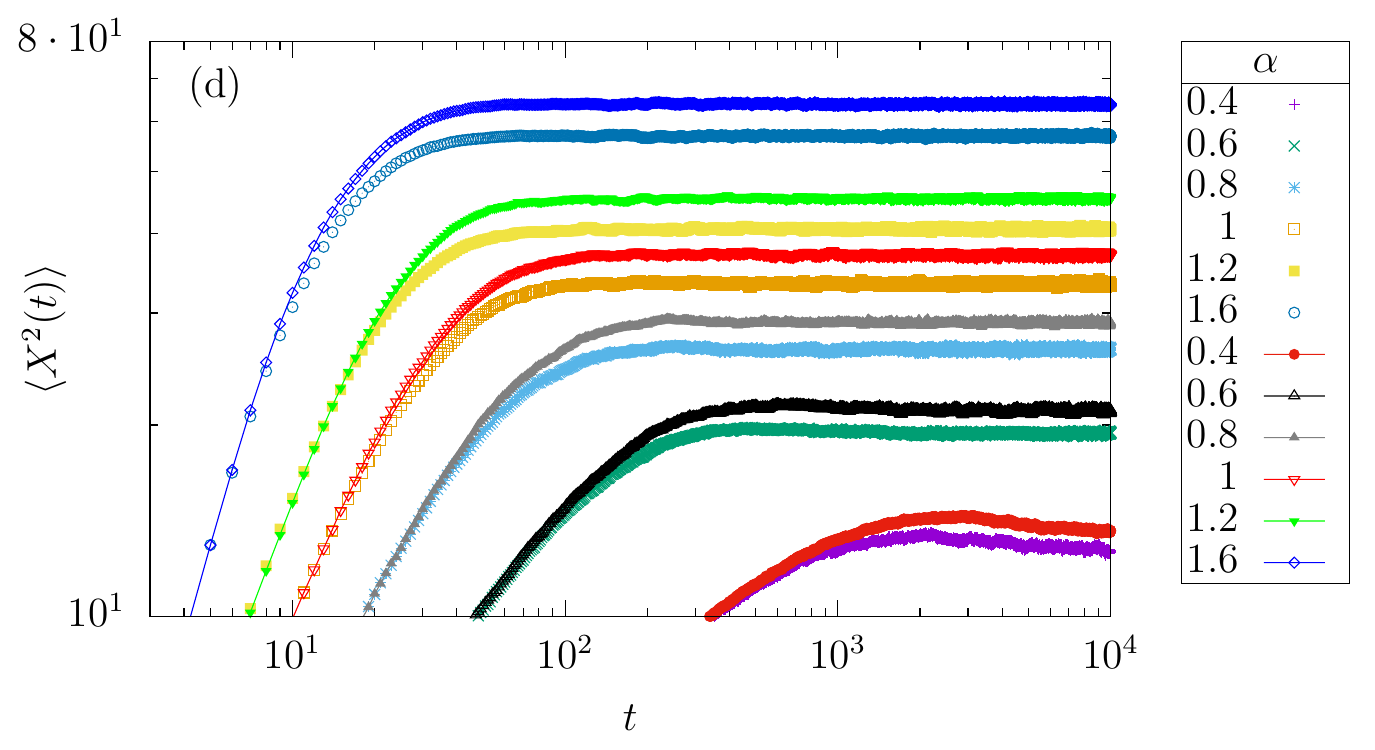}
\includegraphics[width=6.8cm]{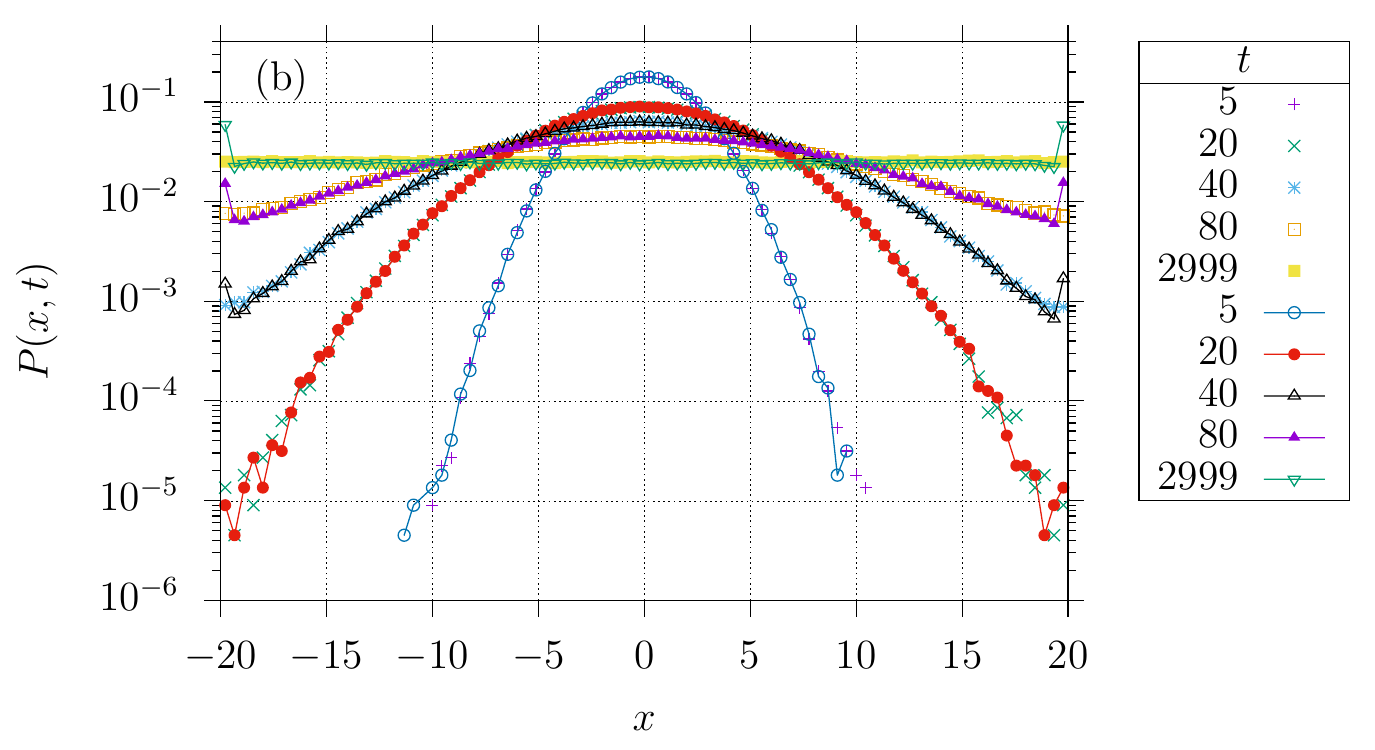}
\includegraphics[width=6.8cm]{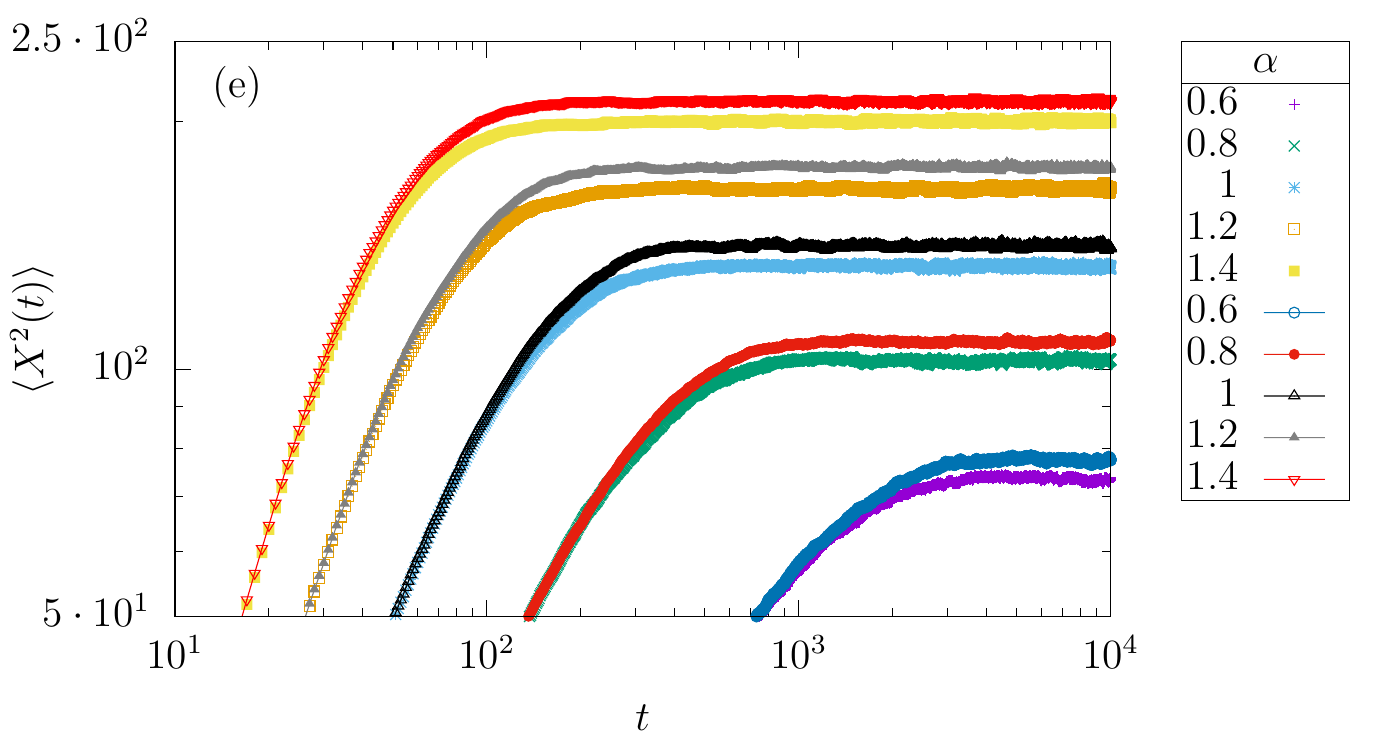}
\includegraphics[width=6.8cm]{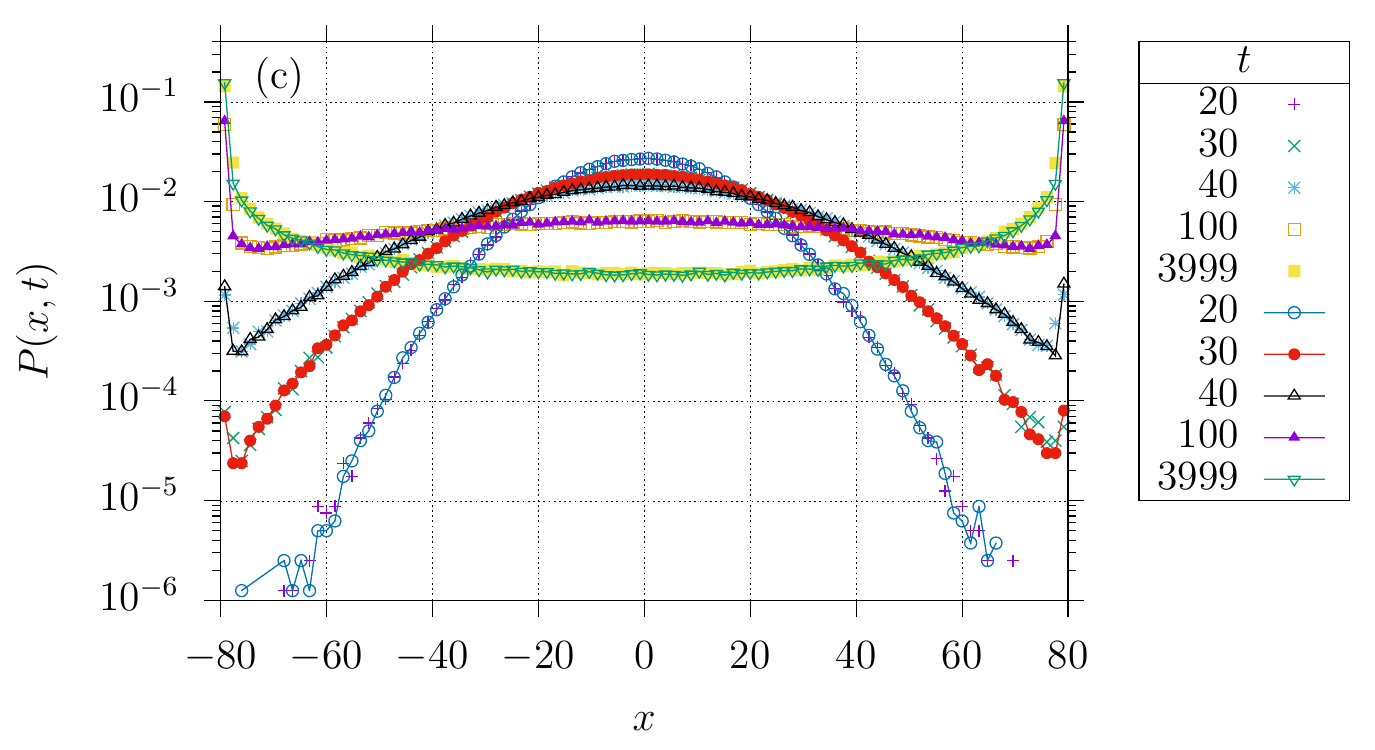}
\includegraphics[width=6.8cm]{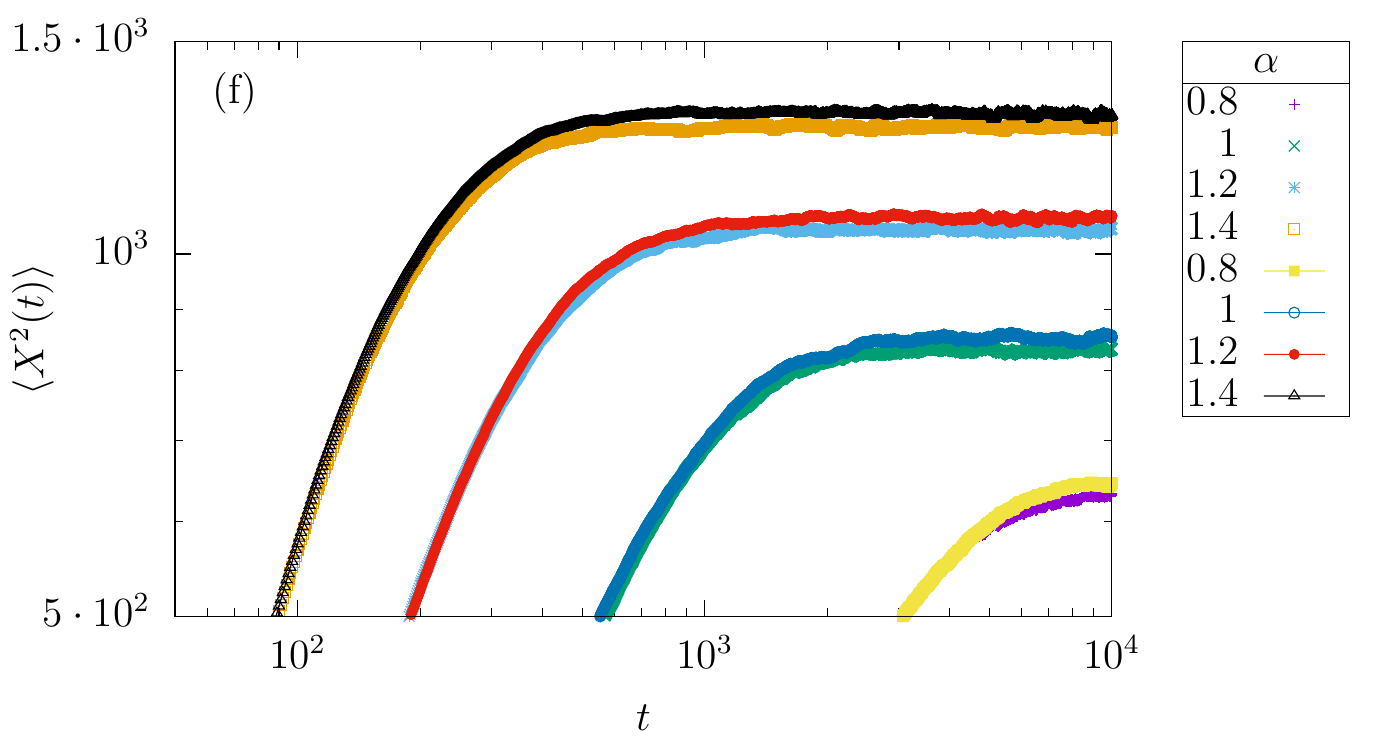}
\caption{\label{fig6}Left panels: Probability density of reflected (points) and
condensed reflected (lines and points) FBM at different times for different
anomalous diffusion exponents, from top to bottom: $\alpha=0.6$, $1.0$, and $1.8$
(for $L=40$, $40$, and $160$, respectively). Right panels: MSD of reflected (points)
and condensed reflected (lines and points) FBM for different interval lengths, from
top to bottom: $L=20$, $40$, $100$. respectively.}
\end{figure}

In the left panels of figure \ref{fig6} we show the probability density of reflected
and condensed reflected FBM for different anomalous diffusion exponents. At short
times the behaviour for both boundary conditions coincide. This is due to the fact
that boundary effects become significant only at longer times, when a significant
portion of particles
had enough time to reach the boundary. At longer times the probability density of
condensed reflected FBM shows a distinct cusp at the boundary, where the values of
the probability density are larger than those for reflected FBM. This cusp persists
also in the stationary distribution, and is present for normal Brownian diffusion,
as well. The cusp naturally stems from the "condensation" of all reflected particles
right at the position of the wall. Clearly, this boundary condition is not consistent
with the known results for Brownian diffusion, and we therefore discard this
alternative definition. To complement this claim, in the right panels of figure
\ref{fig6} we show the MSD of reflected and condensed reflected FBM for different
interval lengths. At short times the MSD for both boundary conditions coincide. At
longer times they deviate, the stationary value for condensed reflected FBM being
larger than that for reflected FBM. This is due to the fact that the condensed
particles at the boundaries contribute to the MSD with a higher amplitude than the
distributed reflected particles in the proper reflected FBM process. Of course,
the deviations due to the exact choice of how to implement reflecting boundary
conditions is expected to become increasingly less relevant when the size $L$
of the interval becomes larger, $L\gg K_{\alpha}^{1/2}$ at unit time.

\begin{figure}
\centering
\includegraphics[width=8cm]{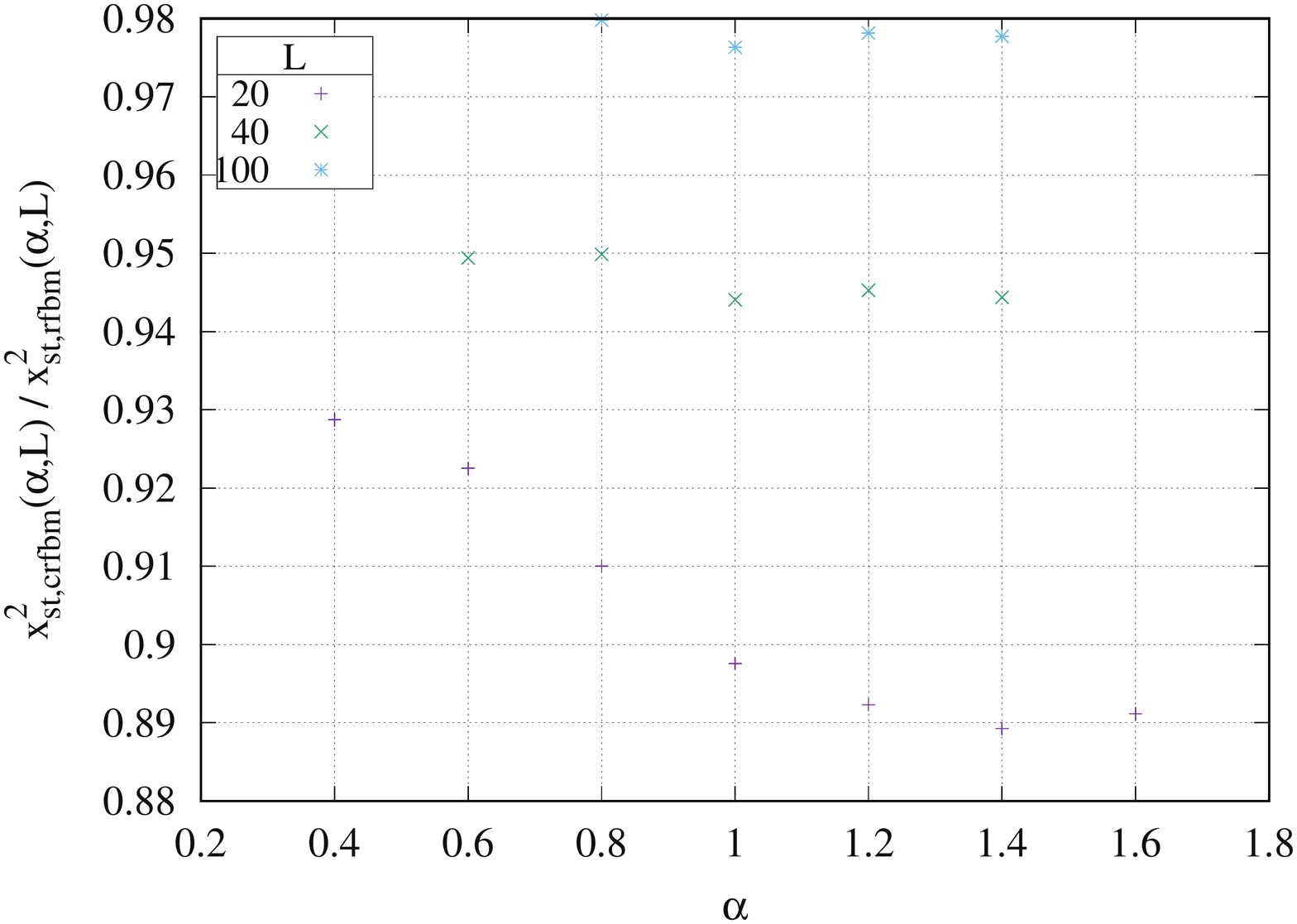}
\caption{Ratio of the stationary MSDs for condensed reflected
FBM and reflected FBM as function of $\alpha$, and for three different interval
lengths $L$.}
\label{ratio}
\end{figure}

Figure \ref{ratio} shows the ratio $x^2_{\mathrm{st,crfbm}}(\alpha,
L)/x^2_{\mathrm{st,rfbm}}(\alpha,L)$ of the stationary MSDs for condensed reflected
FBM and reflected FBM. For the largest interval length, $L=100$, the ratio becomes
practically unity, demonstrating that the definition of the reflecting boundary
condition only disturbs a finite zone around the boundaries.

\section{Stationary mean squared displacement versus $\alpha$}

\begin{figure}
\centering
\includegraphics[width=8cm]{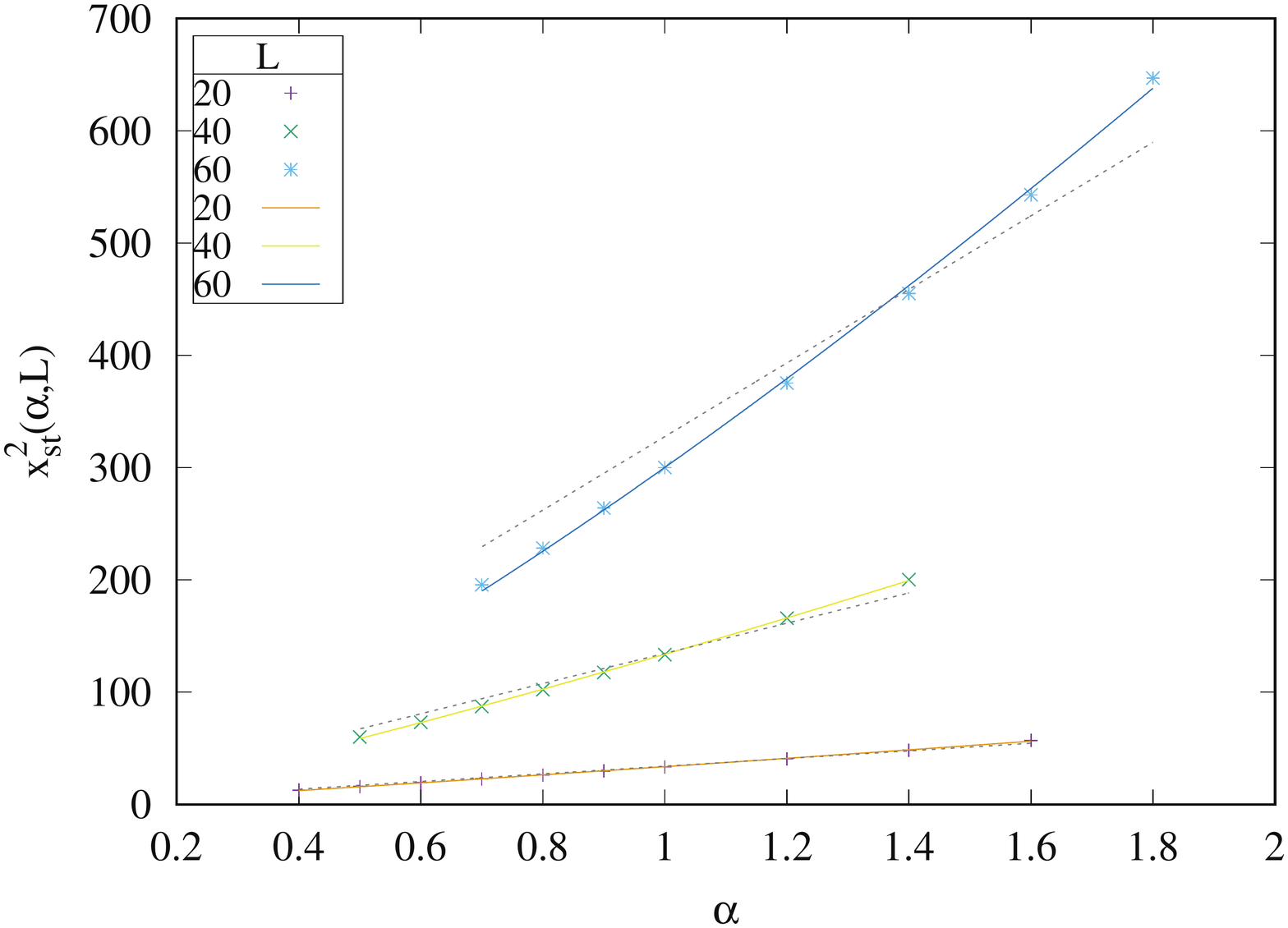}
\caption{Stationary MSD $x^2_{\mathrm{st}}$ as function of $\alpha$.
See text for details.}
\label{x2_vs_alpha}
\end{figure}

Figure \ref{x2_vs_alpha} shows the dependence of the stationary MSD $x^2_{\mathrm
{st}}(\alpha,L)$ on the anomalous diffusion exponent $\alpha$. The dashed lines
show a linear fit while the full lines represent fits to the quadratic form
$g(\alpha,L)=a\alpha^c\times L^2/12$ with the fit parameters $a$ and $c$ listed
in table \ref{tab2}. The fact that $a\approx1$ in all cases corroborates the
$L^2/12$-prefactor in the functional form $x^2_{\mathrm{st}}(\alpha,L)=\alpha^c
\times L^2/12$, while $c$ clearly increases with $\alpha$.

\begin{table}
\centering
\begin{tabular}{|c|c|c|}
\hline
$L$ & $a$  & $c$\\\hline
20  & 1.01 & 1.10\\
40  & 1.00 & 1.19\\
60  & 1.00 & 1.28\\\hline
\end{tabular}
\caption{Fit parameters for the data of figure \ref{x2_vs_alpha}.}
\label{tab2}
\end{table}

\end{document}